\documentclass{revtex4}
\usepackage[latin9]{inputenc}
\usepackage{amsmath}
\usepackage{amssymb}
\usepackage{graphicx}

\makeatletter

\providecommand{\tabularnewline}{\\}

\@ifundefined{textcolor}{}
{%
 \definecolor{BLACK}{gray}{0}
 \definecolor{WHITE}{gray}{1}
 \definecolor{RED}{rgb}{1,0,0}
 \definecolor{GREEN}{rgb}{0,1,0}
 \definecolor{BLUE}{rgb}{0,0,1}
 \definecolor{CYAN}{cmyk}{1,0,0,0}
 \definecolor{MAGENTA}{cmyk}{0,1,0,0}
 \definecolor{YELLOW}{cmyk}{0,0,1,0}
}

\makeatother

\begin{document}

\title{Solitons and entanglement in the double sine-Gordon model}

\author{A. Alonso Izquierdo}

\affiliation{Departamento de Matemática Aplicada y IUFFyM, Universidad de Salamanca,
Spain}

\author{J. Mateos Guilarte}

\affiliation{Departamento de Física Fundamental y IUFFyM, Universidad de Salamanca,
Spain}

\author{N. G. de Almeida}

\affiliation{Instituto de Física, Universidade Federal de Goiás, 74001-970, Goiânia
GO, Brazil}
\begin{abstract}
The bipartite ground state entanglement in a finite linear harmonic
chain of particles is numerically investigated. The particles are
subjected to an external on-site periodic potential belonging to a
family parametrized by the unit interval encompassing the sine-Gordon
potential at both ends of the interval. Strong correspondences between
the soliton entanglement entropy and the kink energy distribution
profile as functions of the sub-chain length are found. 
\end{abstract}
\maketitle
\textbf{Introduction}. - Entanglement in systems displaying solitonic
solutions has been studied for some specific models \cite{Knoll 01,Lewenstein09,Carr09,Lee05,Partner13}.
Recently, solitons in discrete chains presenting bounded internal
modes have received increasing attention from both experimental and
theoretical point of view \cite{Landa10,Landa13,Milenz13,Schneider12,Landa14,Ulm13,Ejtemaee13,Pyka13}.
Since solitonic solutions of nonlinear models are characterized by
being localized and topologically protected, investigation focusing
on their use for quantum information tasks are in progress. As for
example, in \cite{Marcovitch08} the authors studied the behavior
of entanglement in a chain of particles using the Frenkel-Kontorova
model \cite{FKmodel,FKmodelB,FK model 2}, and they have investigated
the possibility to use solitons as the carrier of quantum information.
In this model the particles of the 1-D chain interact by means of
an elastic force with the nearest neighbors and are subjected to an
on-site (substrate) sine-Gordon potential $V$. The strength of the
elastic couplings $g$ between adjacent particles becomes a parameter
in the model which rules the width (proportional to $\sqrt{g}$) of
the unique soliton energy density lump. The influence of the sG soliton
on the bipartite entanglement was analyzed by means of the von Neumann
entropy, where one of two sub-blocks which conform the chain is always
centered at the kink energy density maximum (kink center). The authors
study the dependence of the von Neumann entropy on the size of the
subblocks for several values of $g$, i.e., for several values of
the soliton width. The result is that the presence of the soliton
enhances the entanglement reaching a maximum when the size of the
sub-block coincides approximately with the soliton width. The investigation
of this issue in interacting multi-soliton solutions is achieved in
\cite{Fujii08}.

Solitons and/or kinks are extended states whose energy densities are
spatially localized in bounded regions in contrast with vacuum energy
densities which are spatially homogeneous. In scalar field theory
on the real line the existence of kinks comes from quantization of
solitary wave solutions of the classical field equations. These traveling
waves have energy densities showing one or several lumps of energy
moving together with their center of mass. Thus, the kink center and
the centers and widths of these lumps characterize the extended state,
see e.g. \cite{Book Rajaranan}. We shall address a system in this
class determined by the family of double sine-Gordon potential energy
densities: 
\[
V(\phi;a)=1-(1-a)\cos(2\pi\phi)-a\cos(4\pi\phi)\hspace{0.6cm},\hspace{0.6cm}a\in[0,1]\quad.
\]
The family parameter $a$ is chosen to be in the unit interval in
order to have a function $V$ semi-definite positive. At $a=0$ the
standard sine-Gordon (sG) model is included, whereas at $a=1$ the
sG model reappears although the argument is twice the standard sine-Gordon
angle. Crucial to us in this work is the existence of kink traveling
waves in any member of this family of models. The structure of the
kink profiles, however, differs with varying values of $a$. In particular,
in the range $0\leq a\leq\frac{1}{5}$ the kink energy density shows
a single maximum at the kink center, i.e., it is formed by only one
lump of energy density. In the range $\frac{1}{5}<a<1$, however,
the kink energy density presents two separated maxima symmetrically
distributed with respect to the kink center. The kink profile is formed
by two separated energy density lumps if $a$ is greater than $\frac{1}{5}$
and less than $1$. The distance between the lump centers grows with
$a$ and becomes infinity at $a=1$ where the one-lump sine-Gordon
kink of double angle emerges, see References \cite{Alberto11,Campbell86}.

We shall work in the Frenkel-Kontorova scheme choosing as on-site
substrate the potential of the double sine-Gordon model, i.e., we
shall focus on the double sine-Gordon model defined on a finite chain
of points instead of on the whole real line. Following the ideas disclosed
in \cite{Marcovitch08}, we shall study soliton entanglement entropy
replacing the standard sine-Gordon on-site potential by one potential
belonging to the double sine-Gordon (dSG) family. By this token we
enlarge the number of physical parameters. Besides the spring constant
$g$, the family parameter $a\in[0,1]$ plays a significant r$\hat{{\rm o}}$le.
The length where the kink energy density departs from the vacuum energy,
the kink width, is proportional to $\sqrt{g}$ whereas $a$ determines
the kink profile shape. These facts together make the dSG model specially
suitable to investigate entanglement entropy when the kink energy
distributions are more or less extended over the chain and/or their
shape exhibits one or two energy lumps with different separations.
With this in mind we divide the chain following \cite{Marcovitch08}
into two complementary sub-blocks or sub-chains with no common points.
Unlike Marcovitz et al. we select the two sub-chain starting from
the two chain endpoints. We expect that the entanglement between fluctuations
in different sub-chains will be influenced by how much of the kink
energy distribution is contained in each sub-chain. The $g$ parameter
leads us to distinguish three characteristic regimes: (1) First, we
address the situation where the kink width is much shorter than the
total chain length. For instance, the choice of $g=10^{4}$ on a chain
of length $501$ points (used in this paper) the kink energy density
occupies one fifth of the chain. The kink is clearly localized. (2)
Second, in a chain of the same length a coupling of $g=10^{5}$ concentrates
the kink in three fifths of the chain. The kink is more spread throughout
the chain in this regime than in the previous one. (3) Third, choosing
again the same chain but having $g=10^{6}$ we encounter the last
regime: the kink energy distribution encompasses the whole chain.
Clearly, similar regimes for longer chains are found for higher values
of $g$. In all these considerations it is assumed that the kink center
sits at the chain midpoint.

We shall perform numerical calculations of the correlation functions
and the von Neumann entropy in these three regimes. Both the quantum
vacuum and kink sectors will be investigated, whereas three values
of the $a$ parameter will be considered distinguishing kink profiles
with a single lump, two close lumps, and two widely separated lumps.
Our results offer a refinement in several directions on the structure
of the soliton entanglement entropy discussed in \cite{Marcovitch08}.
First, the spatial correlation functions computed before the partition
into two subchains show shapes in strong correspondence with the kink
energy distribution profile. These correspondences are unveiled in
the three regimes previously described, in each of them for the three
characteristic values of $a$. Comparison with spatial correlation
between vacuum fluctuations is also offered. In a second step, we
use the knowledge of the correlation functions to compute, also numerically,
the entanglement entropy arising in a partition of the system into
two subchains between fluctuations living in different parts of the
chain. Our $N$-particle chain will thus be divided into one $\ell$-particle
and another $N-\ell$ particle subblock with all the $\ell$ points
in the first subchain located at the left of the $N-\ell$ points
of the second subchain. We shall compute the entanglement for different
choices of the partition length $\ell$. The main conclusion is that
entanglement between fluctuations in the two different subchains is
maximum when $\ell$ coincides with one of the kink lump centers.
If the kink energy density shows two energy lumps the entanglement
entropy as a function of $\ell$ reaches two maxima too precisely
when the length of the subblock $\ell$ coincides with the kink lump
centers. In more general terms, we claim that the entanglement is
enhanced when the kink energy density at the sub-block endpoint increases
and it is maximum when the kink energy density is maximum.

\vspace{0.3cm}

\textbf{Quantization}. - We shall address a 1-D linear harmonic chain
of $N$ particles interacting with the nearest neighbors and subjected
to an on-site (substrate) potential $V$. We assume that the equilibrium
position of the $n$-th particle in the chain is $X_{n}^{V}=n$, $n=0,\dots,N$.
The classical Hamiltonian governing the dynamics of this model is
\begin{equation}
H=\sum_{n=0}^{N-1}\,{\cal H}_{n,n+1}=\sum_{n=0}^{N-1}\frac{1}{2}\left[\pi_{n}^{2}+2V(\phi_{n})+g(\phi_{n+1}-\phi_{n})^{2}\right].\label{H1}
\end{equation}
where $\phi_{n}$ represents the displacement of the $n$-th particle
with respect to its equilibrium position, $X_{n}=n+\phi_{n}$, $\pi_{n}=\dot{\phi}_{n}$
is the momentum of the $n$-th particle, $g$ is the non dimensional
strength of the elastic couplings between adjacent neighbor particles,
and $V$ represents the external on-site potential, see \cite{FKmodel}.
We assume that $V(\phi)$ is a non negative function which vanishes
when the displacements reach an equilibrium point. In formula (\ref{H1})
all the magnitudes involved, particle positions, momenta, and potential
are rescaled to non dimensional quantities. The system of coupled
ODE's 
\begin{equation}
\frac{d^{2}\phi_{n}}{dt^{2}}=-\frac{\partial V(\phi_{n})}{\partial\phi_{n}}+g(\phi_{n+1}-2\phi_{n}+\phi_{n-1})\,\,\,,\hspace{1cm}n=0,1,\dots,N\label{odes}
\end{equation}
are the equations of motion describing the classical dynamics of the
model, whereas time-independent solutions are determined from the
difference equations system: 
\begin{equation}
-\frac{\partial V(\phi_{n})}{\partial\phi_{n}}+g(\phi_{n+1}-2\phi_{n}+\phi_{n-1})=0\hspace{1cm}.\label{eq:Diff}
\end{equation}
We shall distinguish two types of time-independent finite energy solutions: 
\begin{enumerate}
\item Homogeneous solutions of (\ref{eq:Diff}), denoted as $\phi^{{\rm V}}(n)=\phi_{n}^{{\rm V}}$:
\[
\phi^{{\rm V}}(n)=\phi_{n}^{{\rm V}}=0\hspace{0.5cm}\mbox{with}\hspace{0.5cm}n=0,1,\cdots,N\,\,\,.
\]
All the particles sit at their equilibrium points $X_{n}=n$ in such
a way that their classical energy is zero: $H(\phi_{n}^{{\rm V}})=0$.
We shall refer to them as \textit{vacuum} solutions foreseeing future
treatment in a quantum setting. 
\item Discrete soliton/kink solutions, which we shall denote as: $\phi_{n}^{S}=\phi^{S}(n)$.
Soliton and kink are frequently used as synonymous in the Literature.
The strict concept of soliton, however, is more restrictive. Both
kinks and solitons are solitary non-linear waves, but they differ
in the fact that solitons preserve shapes after collisions. We shall
denote as \textit{kinks} the non-homogeneous solutions from now on
throughout this paper. Kinks are thus solutions of (\ref{eq:Diff})
obeying the following boundary conditions: 
\begin{equation}
\phi^{S}(0)=\phi_{0}^{S}=0\hspace{1cm},\hspace{1cm}\phi^{S}(N)=\phi_{N}^{S}=1\,\,,\label{eq:contour2}
\end{equation}
characterized accordingly by the non null \lq\lq topological charge\rq\rq:
$T=|\phi_{N}^{S}-\phi_{0}^{S}|=1$. The boundary conditions (\ref{eq:contour2}),
tantamount to $X_{0}^{S}=0$ and $X_{N}^{S}=N+1$, demand that the
kink profile will depart from the particle equilibrium positions somewhere
in the middle of the chain. The kink energy distribution (density
in the continuum limit) is: ${\cal E}(\phi_{n}^{S})=V(\phi_{n}^{S})+\frac{1}{2}g(\phi_{n+1}^{S}-\phi_{n}^{S})^{2}$. 
\end{enumerate}
Small fluctuations $\eta(n,t)$ around static solutions $\phi^{{\rm sta}}(n)=\phi_{n}^{{\rm sta}}$,
either vacuum or kink configurations, 
\[
\phi(n,t)=\phi_{n}^{{\rm sta}}+\eta(n,t)
\]
are still solutions of the ODE's system (\ref{odes}) if and only
if the following ODE's linear system 
\begin{equation}
\Big(\frac{d^{2}}{dt^{2}}+\Omega(n)+2g\Big)\eta(n,t)-g\,\eta(n-1,t)-g\,\eta(n+1,t)={\cal O}(\eta^{2})\label{lodes}
\end{equation}
is satisfied, see \cite{Alberto11}. In formula (\ref{lodes}) $\Omega$
is the second derivative of the potential evaluated at the static
solution: \hspace{1cm}$\Omega(n)=\frac{d^{2}V}{d\phi^{2}}\left[\phi_{n}^{{\rm sta}}\right]$.

If $\phi_{n}^{{\rm sta}}=\phi_{n}^{V}$ then $\Omega(n)$ is the curvature
of the potential at the equilibrium points, where all the particles
of the chain sit. If $\phi_{n}^{{\rm sta}}=\phi_{n}^{S}$ then $\Omega(n)$
is evaluated at kink solutions and because kinks are not homogeneous,
i.e., they vary with $n$, it takes different values at different
points of the chain. From the spectral problem of the $(N+1)\times(N+1)$
Hessian matrix $B(n,m)$ 
\begin{equation}
B(n,m)=\Big(\Omega+2g\Big)\delta_{nm}-g\,\delta_{n,m-1}-g\,\delta_{n,m+1}\,\,\,,\,\,\,\,\,\sum_{m=0}^{N}B(n,m)\psi_{k}(m)=\omega_{k}^{2}\psi_{k}(n)\label{espectral}
\end{equation}
where $\Omega$ obeys to either to vacuum or kink static solutions,
the general solution of the ODE's system (\ref{lodes}) 
\begin{equation}
\eta(n,t)=\sum_{k=0}^{N}\,\frac{1}{\sqrt{2\omega_{k}}}\left(a_{k}\, e^{i\omega_{k}t}+a_{k}^{*}\, e^{-i\omega_{k}t}\right)\psi_{k}(n)
\end{equation}
is obtained in terms of the $(N+1)$ normal modes of fluctuation $\psi_{k}(n)$,
whose eigenvalues have been denoted as $\omega_{k}^{2}$, see (\ref{espectral}).
Dirichlet boundary conditions ensure that the eigenmodes are real.

Quantization of this pre-quantum picture is achieved by promoting
the $a_{k}$ and $a_{k}^{*}$ coefficients to annihilation and creation
bosonic operators: $[\widehat{a}_{k}^{\dagger},\widehat{a}_{q}]=\delta_{kq}$.
The bosonic quantum field and its conjugate momenta operator read:
\begin{equation}
\widehat{\eta}(n,t)=\sum_{k=0}^{N}\frac{1}{\sqrt{2\omega_{k}}}\,\left(\widehat{a}_{k}\, e^{i\omega_{k}t}+\widehat{a}_{k}^{\dagger}\, e^{-i\omega_{k}t}\right)\psi_{k}(n)\,\,\,,\,\,\,\,\,\widehat{\pi}(n,t)=i\sum_{k=0}^{N}\sqrt{\frac{\omega_{k}}{2}}\,\left(\widehat{a}_{k}\, e^{i\omega_{k}t}-\widehat{a}_{k}^{\dagger}\, e^{-i\omega_{k}t}\right)\psi_{k}(n)\,\,.\label{eq:Pi}
\end{equation}
States characterized by a set of occupation numbers $n_{k}$ of each
$k$-th normal mode are eigenfunctions of the linearized Hamiltonian
$\widehat{H}_{0}=\sum_{k=0}^{N}\,\left(\widehat{a}^{\dagger}(k)\widehat{a}(k)+\frac{1}{2}\right)\omega_{k}$
of energy: 
\begin{equation}
\widehat{H}_{0}\vert\left\{ n_{k}\right\} \rangle=E_{\left\{ n_{k}\right\} }\vert\{n_{k}\}\rangle\quad,\quad E_{\left\{ n_{k}\right\} }=\sum_{k=0}^{N}\left(n_{k}+{\textstyle \frac{1}{2}}\right)\omega_{k}\,\,.\label{quen}
\end{equation}
These Fock space states correspond to the fundamental mesons of the
system, both in the vacuum sector and in the kink sector, distinct
alternatives distinguished by the appropriate choice of the $B$ matrix.

\vspace{0.3cm}

\textbf{The double sine-Gordon model}.- We choose the substrate potential
between the members of the one-parametric family of double sine-Gordon
(dSG) potentials 
\begin{equation}
V(\phi;a)=1-(1-a)\cos(2\pi\phi)-a\cos(4\pi\phi)\,\,\,,\,\,\, a\in[0,1]\,,\label{eq:DSG}
\end{equation}
plotted in Figure 1 as a function of $\phi$ for three characteristic
values: $a=0.0$, $0.6$ and $0.99$, see \cite{Mussardo07}. 
\begin{figure}[h]
\centerline{}\includegraphics{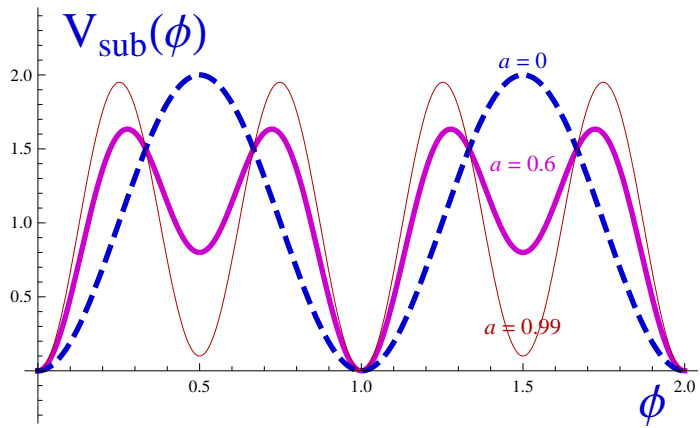}

\caption{Double sine-Gordon potential for the parameters $a=0.0$ (blue dashed
line), $0.6$ (pink thick solid line) and $0.99$ (red thin line). }
\end{figure}

The set of absolute minima or zeroes of $V(\phi;a)$ in the interval
$a\in[0,1)$ is independent of $a$: ${\cal M}=\{\phi^{(j)}=j\,\,,\,\, j\in\mathbb{Z}\}$,
see Figure 2. As a function of $\phi$ (\ref{eq:DSG}), however, exhibits
more critical points. $\frac{dV}{d\phi}(\phi;a)=0$ if and only if:
\begin{equation}
\phi^{(j)}=j\quad,\quad\phi^{(j+\frac{1}{2})}=j+\frac{1}{2}\quad,\quad\phi^{(\pm,\, j)}=\pm\frac{1}{\pi}{\rm arctan}\sqrt{\frac{3a+1}{5a-1}}+j\quad.\label{crpv}
\end{equation}
Evaluation of the second derivative of the potential at the three
types of critical points 
\begin{equation}
\frac{d^{2}V}{d\phi^{2}}(0;a)=4(1+3a)\pi^{2}\hspace{0.4cm},\hspace{0.4cm}\frac{d^{2}V}{d\phi^{2}}({\textstyle \frac{1}{2}};a)=4(5a-1)\pi^{2}\hspace{0.4cm},\hspace{0.4cm}\frac{d^{2}V}{d\phi^{2}}(\phi^{(+,0)};a)=\frac{1-a(2+15a)}{a}\pi^{2}\label{parm}
\end{equation}
tells us that the first type is always formed by minima of $V$, the
second type contains maxima if $a<\frac{1}{5}$ that become relative
minima when $a$ reaches $\frac{1}{5}$ and the last type formed only
by real critical points for $a>\frac{1}{5}$ that are always maxima,
see Figure 1 and 2. Only quantization over absolute minima is safe,
relative minima give rise to false vacua through tunnel effects, and
maxima produce tachyons. We thus choose the state with no fluctuations
over the classical configuration $\phi^{(0)}$ as the vacuum state
of the quantum system {%
\footnote{We could choose another absolute vacuum $\phi^{(j)}$, $j\neq0$,
to quantize but this would lead to a completely equivalent quantum
system.%
}}. Quasi-particles of mass $4(1+3a)\pi^{2}$ arise when the creation
operator $\hat{a}^{\dagger}$ acts on the vacuum state.

\begin{figure}[h]
\begin{tabular}{c}
\includegraphics{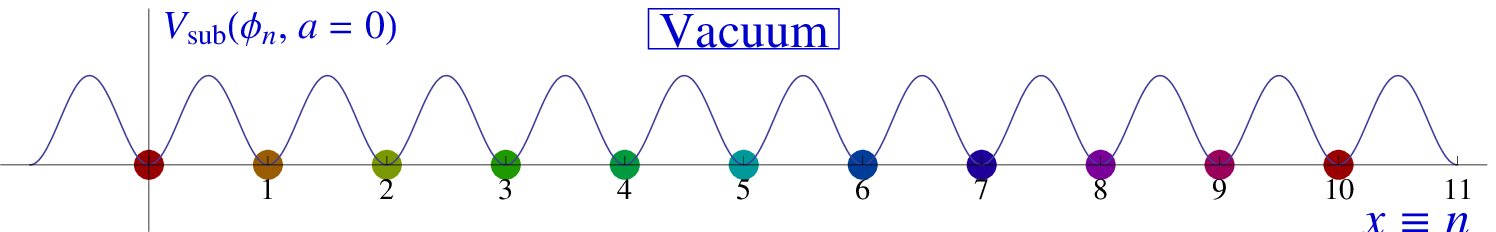}\tabularnewline
\includegraphics{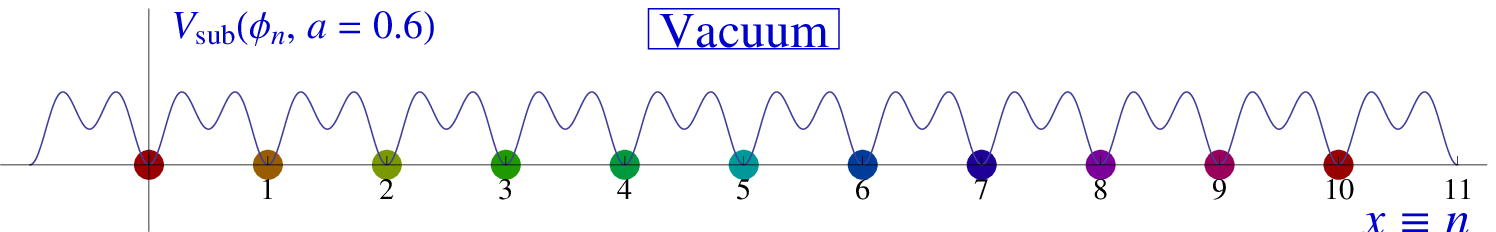}\tabularnewline
\includegraphics{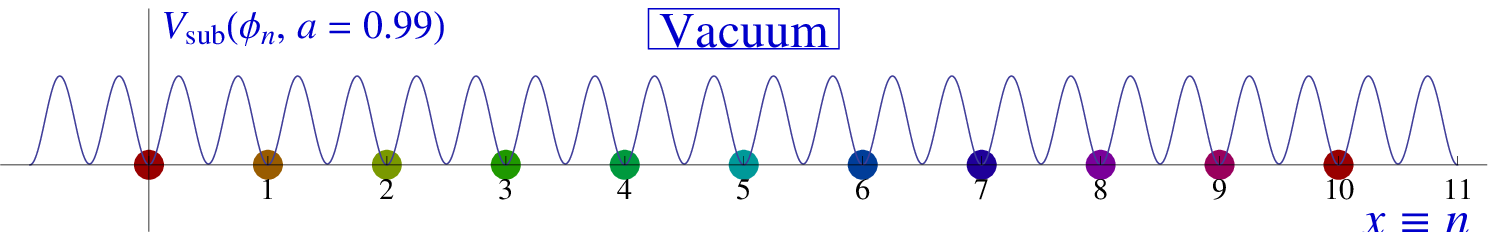}\tabularnewline
\end{tabular}

\caption{Pictures of vacuum solutions for a chain of 11 particles subjected
to a double sine-Gordon substrate potential, corresponding to the
$a$ parameter values: (a) $a=0.0$, (b) $a=0.6$ and (c) $a=0.99$.}
\end{figure}

Besides the homogeneous solutions of the field equations (\ref{eq:Diff})
listed in (\ref{crpv}) there are other static but $n$-dependent
solutions in these model. These are the kinks that must comply with
the contour conditions (\ref{eq:contour2}) at the endpoints because
the system is defined on a finite chain, see Figure 3. 
\begin{figure}[h]
\includegraphics{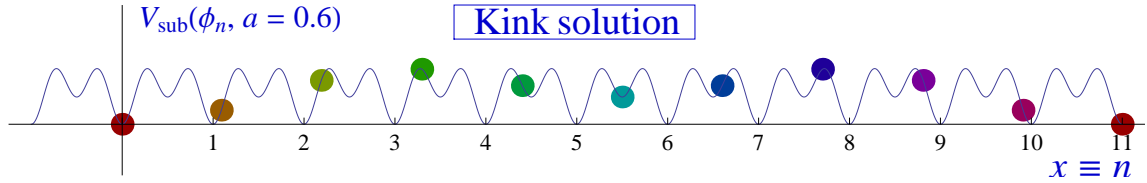}

\caption{Graphics of the kink solution numerically obtained for a chain of
11 particles under the dSG substrate potential and the parameter value:
$a=0.6$. Note that the particle on the right boundary is displaced
in one site with respect to the same particle in the vacuum. The particles
not in the boundary are displaced away from their equilibrium positions. }
\end{figure}

The system of finite differences equations (\ref{eq:Diff}) shows
that these solutions depend both on $a$ and the strength of the elastic
coupling $g$. As stated before the parameter $g$ rules the extension
of this topological solution along the chain, see \cite{Marcovitch08}.
A qualitative analysis of (\ref{eq:Diff}) confirms this claim. If
$g$ is strong enough the particle distribution solving (\ref{eq:Diff})
is such that $\phi_{n+1}-\phi_{n}\simeq\frac{1}{N}$ for almost all
$n$. $\phi(n)$ is almost a straight line with a slope as small as
allowed by the boundary conditions. The influence of the external
potential is very small and the kink profiles depart from the homogeneous
vacuum practically along the whole chain. At weak coupling $g\rightarrow0$,
however, the distribution of particles favoured by kink solutions
of (\ref{eq:Diff}) presents abrupt changes from one to the next particle
somewhere in the middle of the chain. The potential provides strong
forces away the thin kink region restoring the particles to lie on
consecutive absolute minima of $V$ at the two sides of the kink.
In this case the effect of the kink is confined within a small part
of the chain. In Figures 4-5-6 we show graphics of numerically generated
kink solutions with unit topological charge for several values of
the elastic constant $g$ in a chain of $N+1=501$ particles together
with their energy distributions where the above mentioned pattern
is observed. Having fixed this length for the chain we have selected
two values of $g$ in order to characterize the strong and weak regimes
mentioned above plus one third value in between to analyze the intermediate
regime. It is clear that for chains of different length one must adopt
other values of $g$ to deal with kinks that cover $\frac{1}{5}$,
$\frac{3}{5}$, and/or the whole chain. We thus distinguish three
different regimes: 
\begin{enumerate}
\item \textit{Substrate potential dominant regime} (SubsReg). We select
the value $g=10^{4}$ as the regime representative in a $N=500$ particle
chain. The dynamics is fundamentally determined by the substrate potential
in detriment of the elastic inter-particle forces. The kink energy
distribution is localized in an interval of length $\frac{N}{5}=100$. 
\item \textit{Balanced Substrate potential/Elastic force regime} (BalReg).
$g=10^{5}$ is the representative value of the elastic parameter in
a chain of the same length. Elastic forces and those induced by the
substrate potential act on each particle in a balanced manner. Under
these circumstances the kink energy distribution spreads on an interval
of length $3\frac{N}{5}=300$. 
\item \textit{Elastic force dominant regime} (ElasReg). On the same chain
$g=10^{6}$ is a representative value of this third regime. Forces
induced by the substrate potential are very weak as compared with
the elastic interparticle forces. The kink energy distribution is
smoothly distributed over the whole chain. 
\end{enumerate}
\begin{figure}[h]
\centerline{\includegraphics[width=5cm]{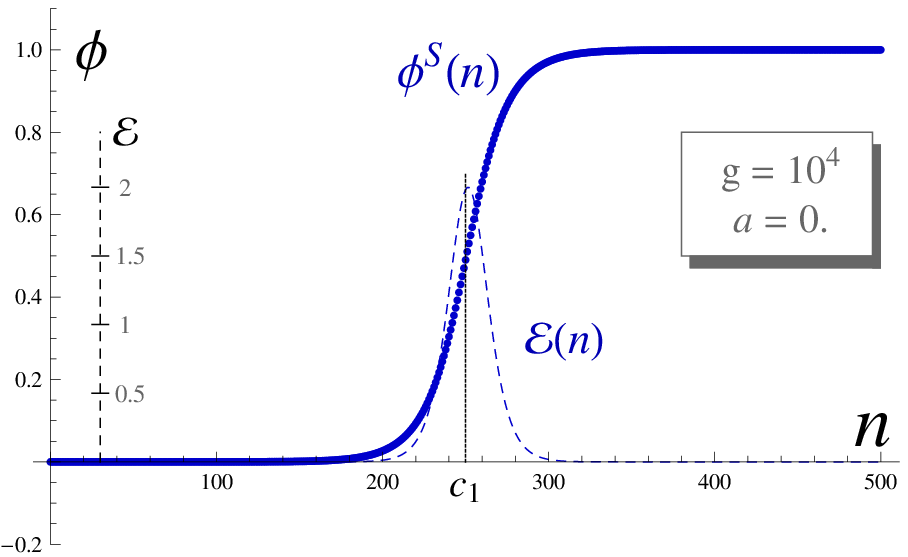}\includegraphics[width=5cm]{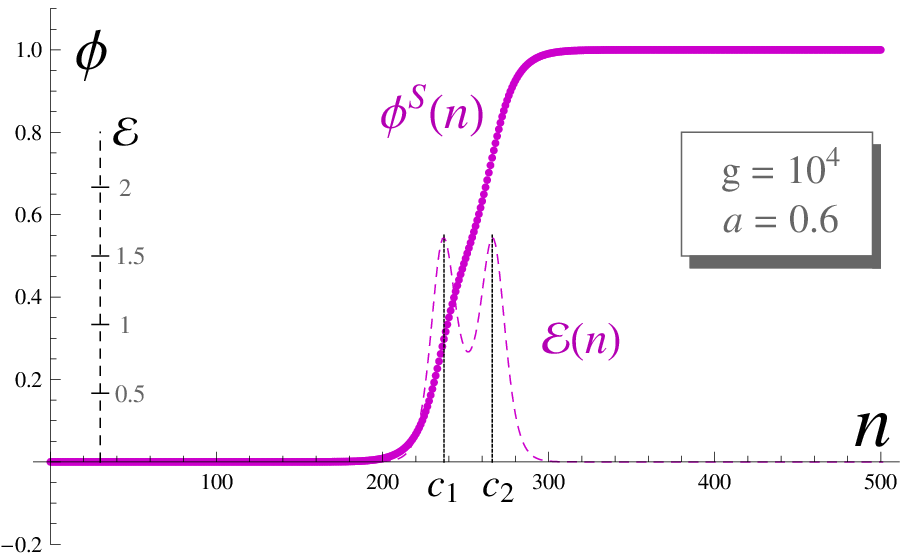}\hspace{0.3cm}
\includegraphics[width=5cm]{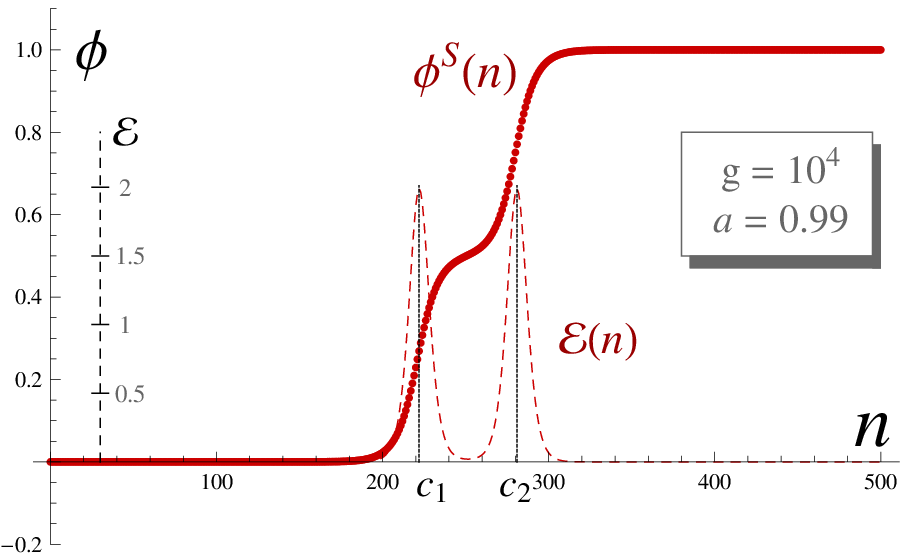}}

\caption{Kink profiles (solid line) and their energy distributions (dashed
line) for $a=0$, $a=0.6$ and $a=0.99$ in the substrate potential
dominant regime of the dSG model on a $N=500$ chain. The kink profiles
jump from a homogeneous solution to the next one in a abrupt manner.
Accordingly, the kink energy distributions neatly depart from zero
only on a short interval and have one maximum if $a<\frac{1}{5}$
but two maxima appear if $a>\frac{1}{5}$. }
\end{figure}

\begin{figure}[h]
\centerline{\includegraphics[width=5cm]{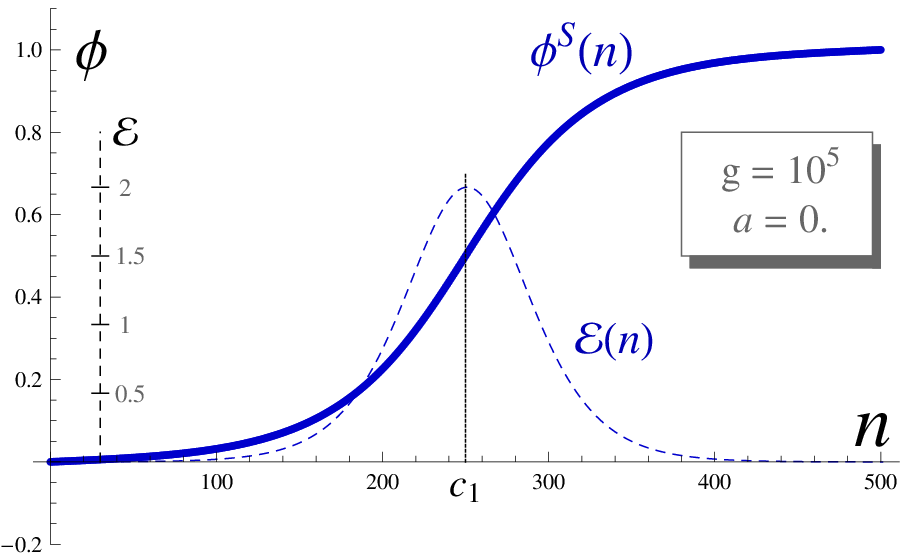}\includegraphics[width=5cm]{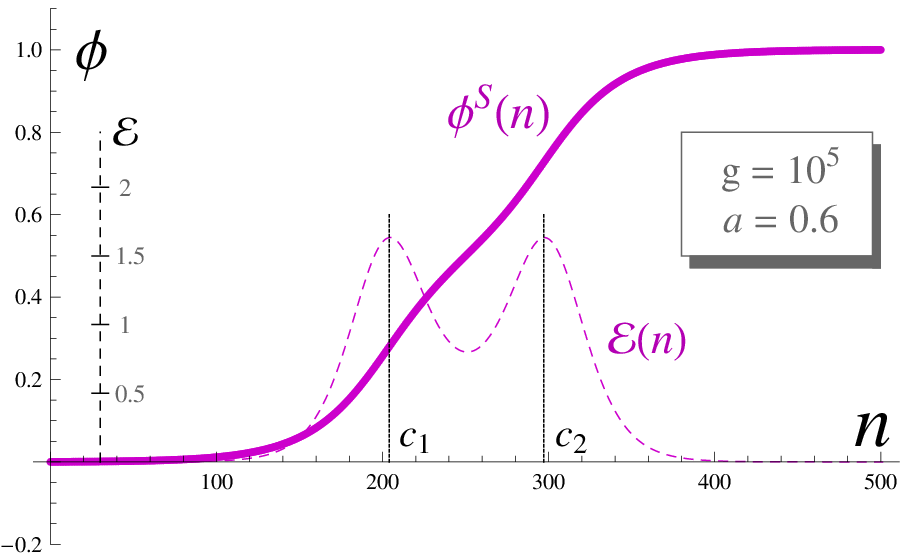}\includegraphics[width=5cm]{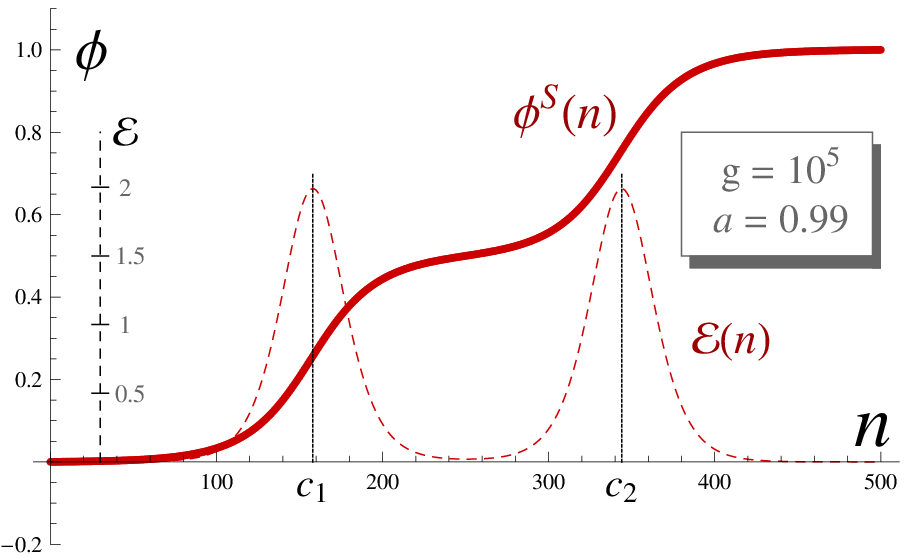}}

\caption{Kink profiles (solid line) and their energy distributions (dashed
line) for $a=0$, $a=0.6$ and $a=0.99$ in the balanced substrate
potential/elastic force regime of the dSG model on a $N=500$ chain.
The jump from a vacuum solution to the next one is less steep than
in the regime discussed in Figure 4. The kink energy distributions
are also less localized and spread over longer intervals. The distinct
number of lumps, one if $a<\frac{1}{5}$, two for $a>\frac{1}{5}$,
is also observed.}
\end{figure}

\begin{figure}[h]
\centerline{\includegraphics[width=5cm]{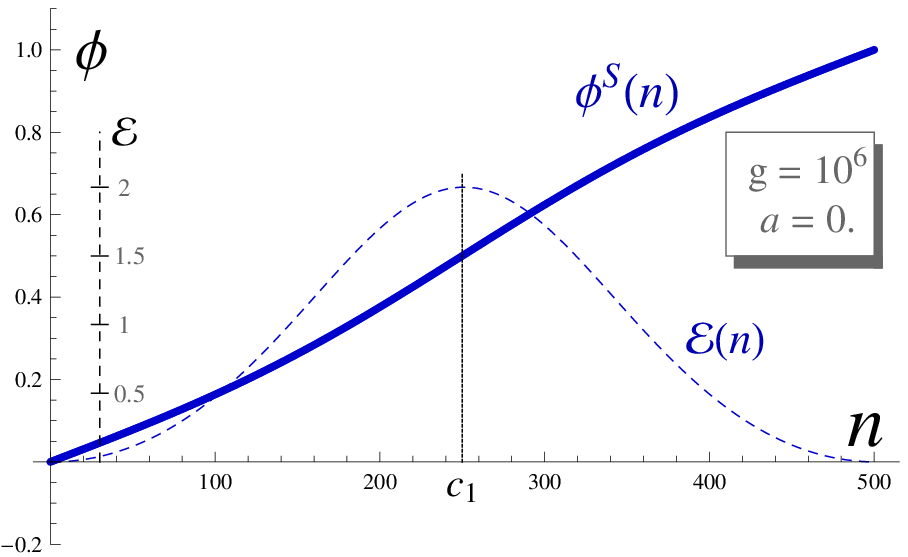}\includegraphics[width=5cm]{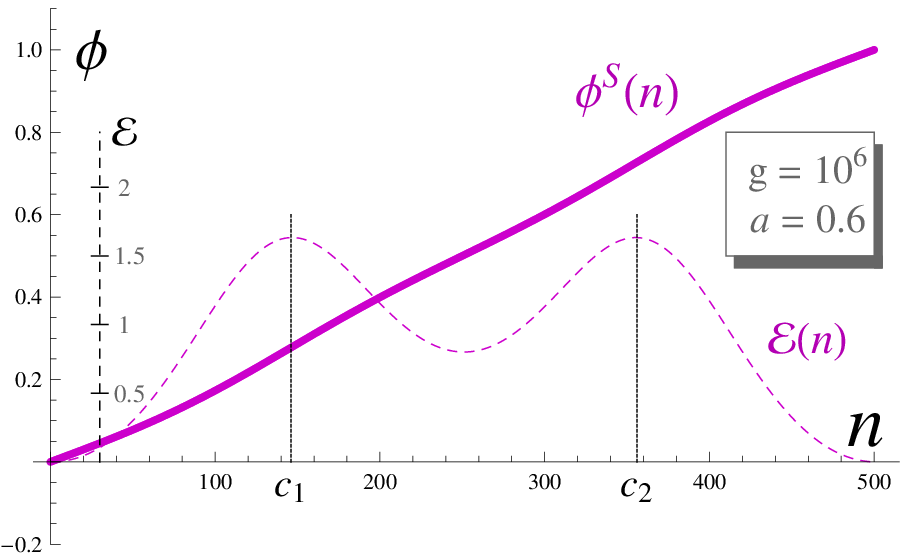}\includegraphics[width=5cm]{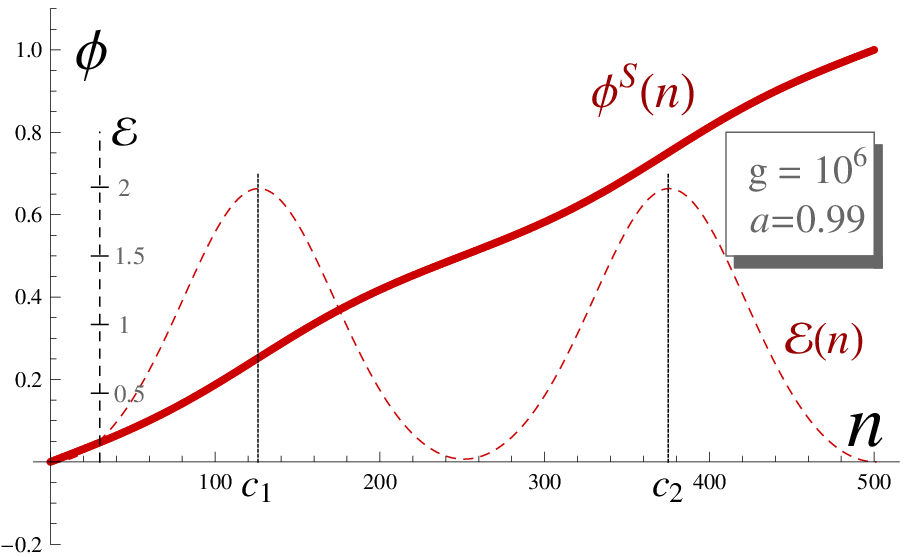}}

\caption{Kink profiles (solid line) and their energy distributions (dashed
line) for $a=0$, $a=0.6$ and $a=0.99$ in the elastic force dominant
regime of the dSG model on a $N=500$ chain. The kink profiles climb
almost uniformly starting from one vacuum solution at the chain left
endpoint and ending at the next vacuum solution on the chain right
endpoint. The kink energy distributions are even less localized than
those in the previously described regime. The existence of energy
distributions formed by one or two lumps when $a$ is below or above
$\frac{1}{5}$ is also appreciated.}
\end{figure}

\newpage{}In sum, the set of graphics displayed in Figures 4-5-6
reveals the r$\hat{{\rm o}}$le of the parameter $a$. For $a$ less
than but close to one the kink profile consists in a two-step climbing,
with a long plateaux in between, from one vacuum solution up to the
next one. The energy distribution is concentrated around the midpoints
of each jump and thus a two energy lump configuration arise. The reason
is the existence of a relative minima of $V$ between each two absolute
minima when $a>\frac{1}{5}$, recall formulas (\ref{crpv}) and (\ref{parm}).
Moreover, because $V(\phi^{{(j+\frac{1}{2})}})=2(1-a)$ the value
of the potential at these relative minima is close to zero when $a$
is slightly less than one. The climb from $\phi^{(j)}$ to $\phi^{(j+1)}$
\lq\lq resting\rq\rq at $\phi^{(j+1/2)}$ as much as possible.
Thus, the distance between the two jumps is very large when, e.g.,
$a=0.99$. When $a$ decreases but is still greater than $\frac{1}{5}$
the situation is the same but the value of the potential at the relative
minimum is higher and the distance between the jumps is shorter, i.e.,
the two lumps in the energy distribution are closer. Finally, if $a<\frac{1}{5}$
the relative minima become maxima and the climbing takes place in
one step because there is no a relative minima in between to rest.
The $a=1$ limit is special. The relative minima become absolute,
zeroes of the substrate potential $V(\phi;1)$, that form the set:
${\cal M}(1)=\{\phi^{{(j)}}={\textstyle \frac{j}{2}}\,\,,\,\, j\in\mathbb{Z}\}$.
For this value the dSG potential (\ref{eq:DSG}) becomes a pure sG
potential again. Certainly the model in the limit $a=1$ involves
a two-kink solution of a single-well sine-Gordon potential with argument
twice the usual angle. In the field theory model on the continuous
real line this limit implies that one of the kinks goes to infinity
because the relative minimum becomes absolute. In a finite discrete
chain, we selected the value $a=0.99$ to offer a description of kink
profiles close to this limit: the two lumps of energy moves towards
the endpoints of the chain.

At this point, it is important to emphasize that the surge of two
lumps on kink profiles when $a>\frac{1}{5}$ in the unit topological
charge sector of the double sine-Gordon does not means that these
soliton structures are multi-soliton or breather solutions akin to
those arising in pure sine-Gordon models, i.e., the models found in
the boundary points $a=0$ and $a=1$ of the family. sine-Gordon multisolitons
live in sectors with topological charge different from one and are
time-dependent even in their center of mass. Unlike in the standard
sine-Gordon model, there exists forces between kinks either formed
by one or two lumps in the double sine-Gordon models, specifically
between kinks and anti-kinks, see \cite{Campbell86}. Isolated kinks,
however, are static in their center of mass: either formed by one
or two lumps the only time-dependence arise from the motion of the
center of mass obeying to Lorentz transformations.

We pass to study how the rich structure of these dSG solitons influences
quantum entanglement between vacuum and kink fluctuations. In particular,
we begin by analyzing the spatial correlation functions between two
particles in the chain in the kink and vacuum sectors.

\vspace{0.2cm}

\textbf{Spatial Correlation Functions}. - To compute spatial correlation
functions in this $N$-particle system we follow the procedure established
in the References \cite{Simon94,Marcovitch08}. The canonical variables
are assembled into a vector $Y=(\eta,\pi)$ which is a point in phase
space. The $N$-component vectors $\eta=(\eta_{1},\eta_{2},...,\eta_{N})$
and $\pi=(\pi_{1},\pi_{2},...,\pi_{N})$ encompass respectively the
particle positions and momenta. Quantization forces the following
commutation rules: 
\begin{equation}
[\widehat{Y}_{\alpha},\widehat{Y}_{\beta}]=iJ_{\alpha\beta}\quad,\quad J=\left(\begin{array}{cc}
\mathbf{0} & \mathbf{1}\\
-\mathbf{1} & \mathbf{0}
\end{array}\right),
\end{equation}
where the $2N\times2N$ symplectic matrix $J$ is written in this
formula as $N\times N$ blocks of the null and identity matrices.
Expectation values of two canonical variables in a quantum state are
the matrix elements of the $2N\times2N$ covariance matrix $M={\rm Re}[\langle\widehat{Y}\widehat{Y}^{T}\rangle]$,
which may be diagonalized in the Williamson form $W=S_{W}MS_{W}^{-1}={\rm diag}\left(\lambda_{1},\lambda_{2},...\lambda_{N},\lambda_{1},\lambda_{2},...\lambda_{N}\right)$,
where the eingenvalues $\lambda_{j}\geq\frac{1}{2}$ are the moduli
of the eigenvalues of the matrix $JM$. Gaussian states are completely
characterized by the covariance matrix $M$ and by their first momenta.
Partitioning the N-mode system into two sets $A$ and $B$, the reduced
covariance matrices $M_{k}={\rm Re}[\langle\widehat{Y}_{k}\widehat{Y}_{k}^{T}\rangle]$,
$k=A,B$, must be put in the Williamson form. Matrix elements of the
covariance matrix $M$ are thus of the form $\left\langle \hat{\eta}(n)\hat{\eta}(m)\right\rangle ={\rm tr}\,\hat{\rho}\hat{\eta}(n)\hat{\eta}(m)$,
$\left\langle \hat{\pi}(m)\hat{\eta}(n)\right\rangle ={\rm tr}\,\hat{\rho}\,\hat{\pi}(m)\hat{\eta}(n)$
and $\left\langle \hat{\pi}(n)\hat{\pi}(m)\right\rangle ={\rm tr}\,\hat{\rho}\,\hat{\pi}(n)\hat{\pi}(m)$,
where $\hat{\rho}$ is the density operator of the quantum state of
the chain and $\hat{\eta}(n)$, $\hat{\pi}(n)$ are the quantum operators
defined in (\ref{eq:Pi}). Assuming that the whole system is in the
ground state we find the correlation functions between particle positions
and momenta 
\begin{equation}
{\rm Re}\left\{ \langle\widehat{\eta}(m)\widehat{\eta}(n)\rangle\right\} =\frac{1}{2}\sum_{l}\frac{\psi_{l}(m)\psi_{l}(n)}{\omega_{l}}\hspace{0.5cm},\hspace{0.5cm}{\rm Re}\left\{ \left\langle \widehat{\pi}(m)\widehat{\pi}(n)\right\rangle \right\} =\frac{1}{2}\sum_{l}\omega_{l}\psi_{l}(m)\psi_{l}(n)\,\,\,,\label{eq:Phi-Phi}
\end{equation}
whereas positions and momenta verify: ${\rm Re}\left\{ \langle\widehat{\pi}(m)\widehat{\eta}(n)\rangle\right\} ={\rm Re}\left\{ \left\langle \widehat{\eta}(m)\widehat{\pi}(n)\right\rangle \right\} =0$.

Correlations between two particle positions are obtained from the
normal modes eigenvalues and eigenfunctions: 
\begin{equation}
\xi_{mn}=\left\langle \widehat{\eta}({\textstyle m})\,\widehat{\eta}(n)\right\rangle =\frac{1}{2}\sum_{k=0}^{N}\frac{\psi_{k}(m)\psi_{k}(n)}{\omega_{k}}\quad.\label{correla}
\end{equation}
In Figures 7 (a)-(b)-(c) the dependence on $n$ of the spatial correlation
function when one of the particles sits at the left endpoint of the
chain $\xi_{0n}=\xi_{n}$ is displayed. Numerical computations relying
on formula (\ref{correla}) have been performed for our chain of $501$
particles, accounting for vacuum and kink normal modes, in the nine
cases where the kink profiles were depicted: three characteristic
values of $a$ and $g$. The properties of correlation functions as
functions of $n$ are clearly observed in the $g=10^{4}$ regime.
Vacuum and kink correlations $\xi_{n}$ do not differ from each other
for small $n$ where the kink profile does not depart too much from
the vacuum solution. At larger $n$ kink fluctuation correlations
remarkably grow bigger than vacuum correlations until reaching one
or two maximum values, precisely at the points where the kink energy
distribution also reaches its maximum values. If $n_{+}$ denotes
the position of the maximum farther away from the origin, kink correlations
start to decrease for $n>n_{+}$. The difference between kink correlations
for $a<\frac{1}{5}$ and $a>\frac{1}{5}$ is that in the first case
there is only a single maximum $n_{0}$ whereas there are two $n_{\pm}$
if $a>\frac{1}{5}$. Kink correlations in the last case decrease after
$n_{-}$, pass a relative minimum at $n_{0}$, and increase again
up to $n_{+}$, see Figure 7(a). We conclude that two-point \textit{particle
spatial correlations in the kink sector recognize somehow the kink
profile shape}. This behavior, albeit less pronounced, is reproduced
in the BalReg regime, Figure 7(b), whereas in the ElasReg regime,
see Figure 7(c), the curvatures at the maxima are very mild and the
correlation functions $\xi_{n}$ for vacuum and kink fluctuations
look similar. We thus conclude that the logarithmic correlations in
the kink sector behave in a remarkable different manner that vacuum
correlations of standard logarithmic class: $\xi_{mn}^{V}=\log\frac{1}{\vert m-n\vert}$.

\begin{figure}[h]
\centerline{\includegraphics[width=4cm]{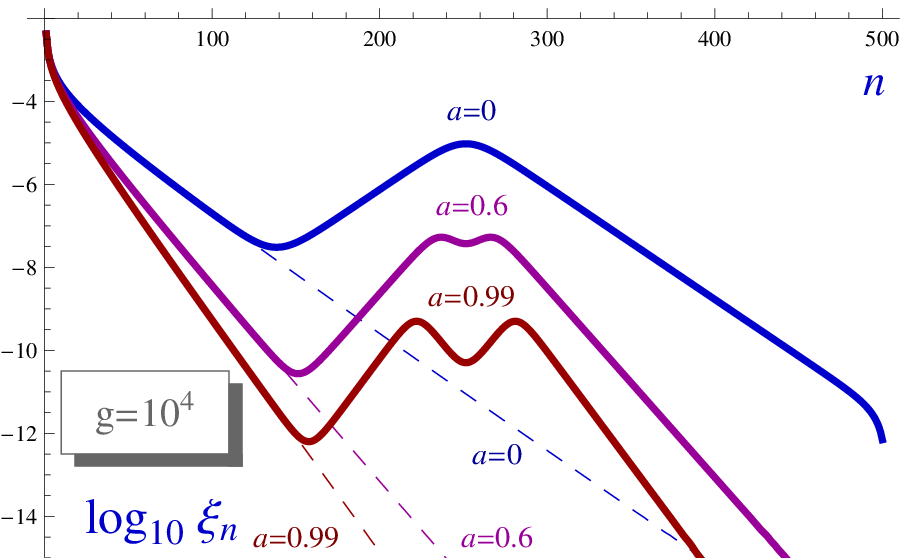}\includegraphics[width=4cm]{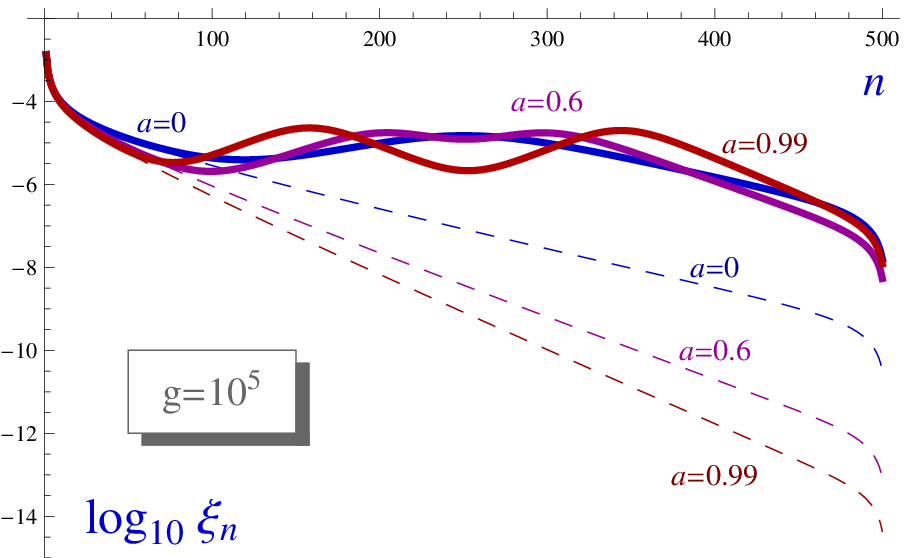}\includegraphics[width=4cm]{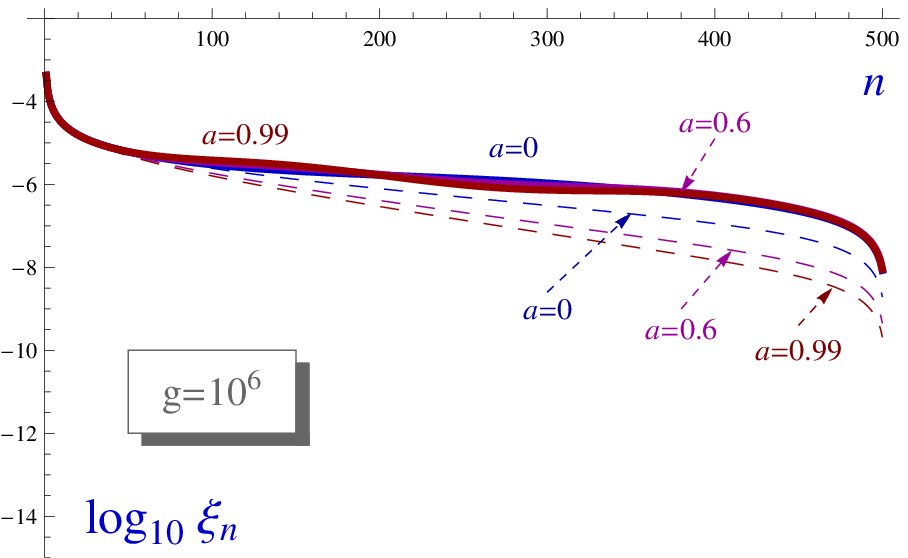}\includegraphics[width=4cm]{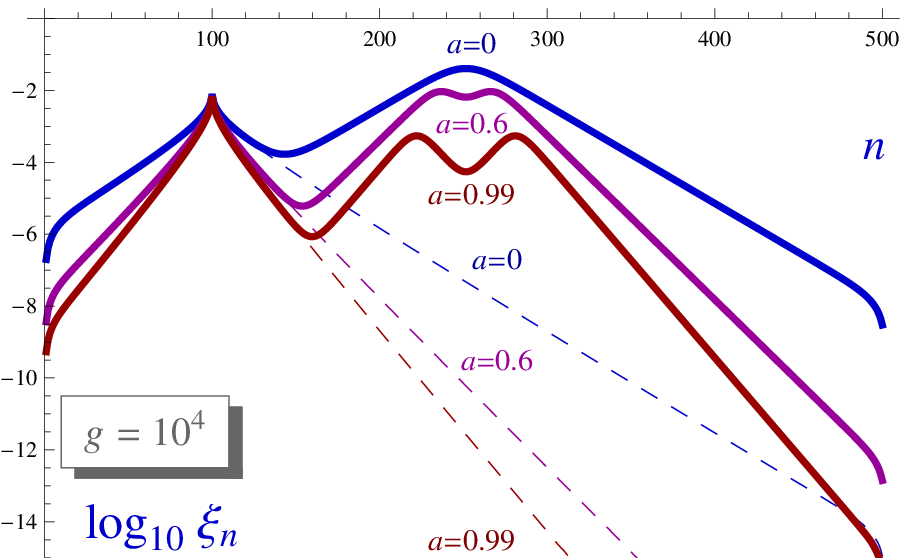}
}

\caption{Graphics of the logarithmic correlation functions between two particles
separated by $n$ positions for vacuum (dashed lines) and kink (solid
lines) fluctuations. Plots of the correlation functions when one of
them is located at the left endpoint of the chain: (a) SubsReg, (b)
BalReg and (c) ElasReg regimes. In each regime, the parameter values
$a=0$, $a=0.6$, and $a=0.99$ have been considered. (d) Graphics
of the logarithmic correlation functions between the $m=100$th particle
and a second one separated by $n$ positions for vacuum (dashed lines)
and kink (solid lines) fluctuations in the $g=10^{4}$ SubsReg regime.
The three characteristic values $a=0$, $a=0.6$, and $a=0.99$ have
been considered.}
\end{figure}

The correlation function shapes as functions of the interparticle
distance just described follows a similar pattern independently of
the fixed particle position. In order to illustrate this fact, correlation
functions between the $m=100$th particle and a second particle located
in an arbitrary point $m+n$ of the $N=500$ chain are displayed in
Figure 7 (d) for $g=10^{4}$. Neither the kink configuration nor the
contour conditions affect the position of the fixed particle in the
correlation function. Recall that the kink profile in this regime
is confined approximately in the range $(200,300)$. We observe that
the correlation function increases when the second particle position
$100+n$ is located at the centers of the kink lumps. In comparison
with the correlations between two particles located respectively at
$0$ and $n$ this effect is reinforced because the fixed particle
is closer to the kink center, see Figures 7a and 7d.

\vspace{0.2cm}

\textbf{Entanglement}. - Entanglement between fluctuations supported
by Gaussian states is measured by taking advantage of the correspondence
between the diagonal form of the covariance matrix $M={\rm Re}\left[\langle YY^{T}\rangle\right]$
for the Gaussian state with the covariance matrix of a thermal state,
also diagonal in Fock space, having an average phonon number $\overline{n}_{j}=\lambda_{j}-\frac{1}{2}$
\cite{Simon94}. Here the $\lambda_{j}$ are the Williamson eigenvalues
of the covariance matrix and entanglement entropy is accordingly defined
as: 
\begin{equation}
E_{S}=\sum_{j}S(\lambda_{j})\hspace{0.5cm}\mbox{with}\hspace{0.5cm}S(\lambda)=\left(\lambda+{\textstyle \frac{1}{2}}\right)\ln\left(\lambda+{\textstyle \frac{1}{2}}\right)-\left(\lambda-{\textstyle \frac{1}{2}}\right)\ln\left(\lambda-{\textstyle \frac{1}{2}}\right).\label{eq:S}
\end{equation}
After choosing a subblock of $\ell$ particles forming a subchain
$A$ of the whole chain of $501$ particles the eigenvalues of the
reduced $2\ell\times2\ell$ covariance matrix to this subblock are
numerically computed. Formula (\ref{eq:S}) is then used to estimate
also numerically the entanglement entropy. In contrast with the usual
choice predominant in the Literature where the subchain is centered
at the chain mid point, see e.g. Reference \cite{Marcovitch08}, we
take the subblock starting from the left endpoint of the chain, see
Figure 8. The reason is that performing these calculations on subchains
of fixed but arbitrary length $\ell\leq501$ a striking similarity
between the entanglement entropy as a function of $\ell$ and the
kink energy distribution along the chain clearly emerges. More precisely:
entanglement between fluctuations in the sub-chains $A$ and $B$
is maximum when $\ell$ is chosen at a maximum of the kink energy
distribution.

In Figures 9-10-11 the entanglement entropies numerically obtained
from (\ref{eq:S}) are displayed corresponding to the three regimes
where the kink profiles were shown before: SubsReg (Figure 9(a)-(b)-(c)),
BalReg (Figure 10(a)-(b)-(c)), and ElasReg (Figure 11(a)-(b)-(c)).
In each regime the graphics for the three values of $a$, $a=0$,
$a=0.6$, and $a=0.99$, are depicted.

\begin{figure}[h]
\centerline{\includegraphics{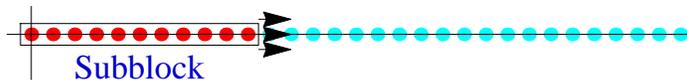}}

\caption{A subblock of particles attached to the left endpoint of the total
chain.}
\end{figure}

Entanglement entropies between vacuum fluctuations depend only slightly
on the subchain location. For fixed $g$ and $a$ they vary with $\ell$
in the same qualitative way as those entropies found for centered
subchains. The differences are only quantitative: the heights of the
plateau of these functions are smaller in this new location of the
subblock. Kink fluctuation entanglement entropies require a much more
detailed analysis. We shall describe in turn the results in the nine
different cases:

\vspace{0.2cm}

A. \textit{Substrate potential dominant regime}

\vspace{0.2cm}

The numerically computed entanglement entropy is displayed in Figures
9(a)-(b)-(c) for the three habitual values of the $a$ parameter.
The kink energy distributions are also shown as a shadowed zone in
the same Figures. Note the correspondence between the maxima of the
kink energy distributions and the entanglement entropy as functions
of $\ell$, in concordance also with the spatial correlation function
maxima.

\begin{figure}[h]
\centerline{\includegraphics[width=5cm]{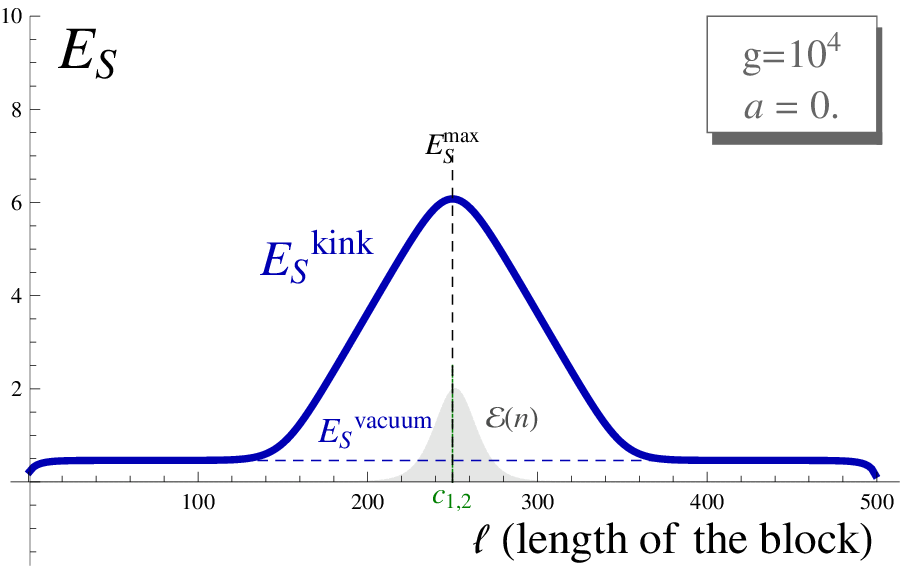}\includegraphics[width=5cm]{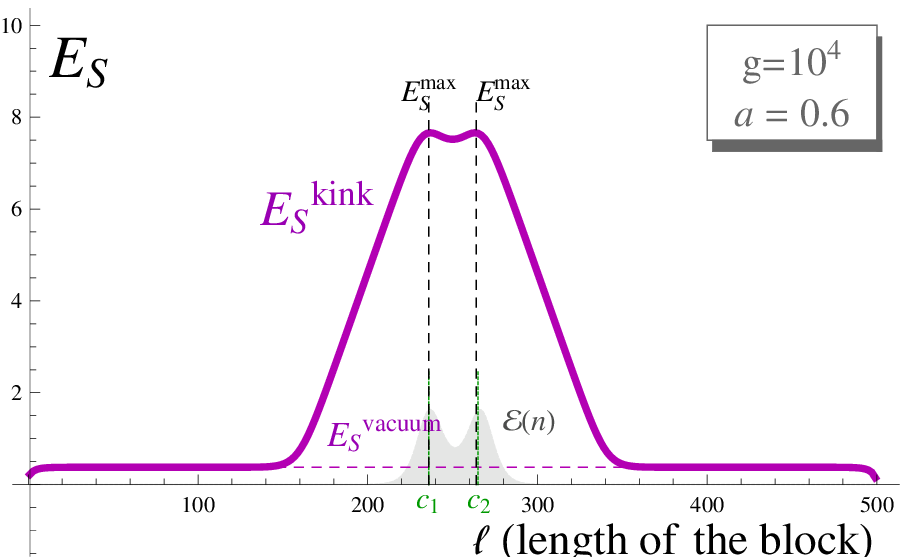}\includegraphics[width=5cm]{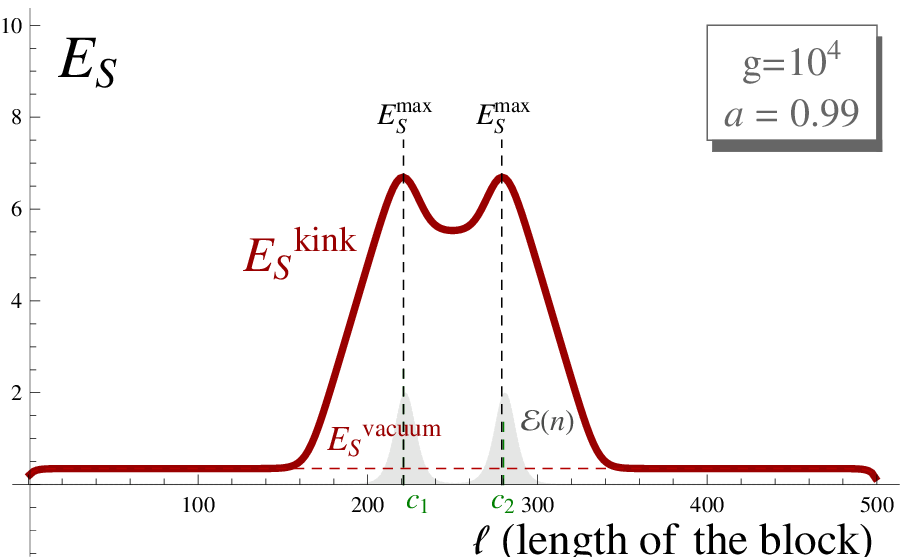}}

\caption{Entanglement entropy between kink-kink (solid line) and vacuum-vacuum
(dashed line) fluctuations for: (a) $a=0$, (b) $a=0.6$ and (c) $a=0.99$
in the SubsReg regime $g=10^{4}$. The kink energy distributions are
plotted as curves enclosing a shadowed area in the scale set by $E_{S}$.
Dashed vertical lines have been included to point out the relative
positions of the kink energy distribution and entanglement entropy
maxima.}
\end{figure}

\vspace{0.2cm}

We remark the following facts: (1) For sine-Gordon kinks, $a=0$,
the entanglement entropy between kink-kink and vacuum-vacuum fluctuations
are identical when the subblock length $\ell$ is small. The kink
supported entropy $E_{S}^{{\rm kink}}$ begins to increase and starts
to differ from $E_{S}^{{\rm vacuum}}$ when $\ell$ approaches 150.
The kink entanglement entropy reaches its maximum value at a subblock
length of $\ell=250$. From this point onwards $E_{S}^{{\rm kink}}$
decreases until a subblock length of approximately $\ell=350$, a
point where again the kink supported entropy equals the vacuum entropy,
see Figure 9(a). The maximum of the entanglement entropy and the center
of the kink energy distribution are identical and lie at the mid point
of the chain, a fact shown by the vertical dashed line. \textit{The
kink entanglement entropy is maximum for a subchain whose endpoint
coincides with the kink energy distribution maximum point}. Again
this is in concordance with the spatial correlation function, which
reaches its maximum just at the center of the single kink, see Figure
7.

(2) When $a=0.6$ the entanglement entropy between kink fluctuations
behaves in a slightly different way. Like in the previous case $E_{S}^{{\rm kink}}$
fits in the function $E_{S}^{{\rm vacuum}}$ approximately until $\ell=150$.
For subblocks of length $150<\ell<236$ the kink fluctuation entropy
increases and reaches a maximum at $\ell=236$, precisely the center
of the first lump of the kink energy distribution. The entropy of
subchains of lengths in the interval $\ell\in(236,250)$ decreases
until hitting a relative minimum $\ell=250$, half the whole chain
length. Beyond this length the entanglement entropy between kink fluctuations
starts to increase again reaching a second maximum at the center of
the second lump of the kink energy distribution $\ell=264$. For subblock
lengths in the interval $264<\ell<350$, $E_{S}^{{\rm kink}}$ decreases
until approximately $\ell=350$. Beyond this subblock length the kink
fluctuation entropy coincides again with $E_{S}^{{\rm vacuum}}$,
see Figure 9(b).

(3) If $a=0.99$ the entanglement entropy $E_{S}^{{\rm kink}}$ follows
the same pattern explained in the previous case, see Figure 9(c).
Here the distance between the two maxima of the entanglement entropy
is greater than the distance between the maxima of the $a=0.6$ kink.
The positions of the two maxima occur in this case for subblock lengths:
$\ell=221$ and $\ell=279$, the maxima of the kink energy distribution.
The kink fluctuation entropy exhibits a relative minimum between the
two maxima but the value of the kink entropy at this minimum is much
greater than $E_{S}^{{\rm vacuum}}$ in the same range.

In the three cases the kink entanglement entropies and energy distributions
are curves as functions of $\ell$ with similar shapes and only the
scales are different. One also checks that the correlations functions
in Figure 7(a)-(d) have the same critical points than the kink energy
distributions.

\vspace{0.2cm}

B. \textit{Balanced Substrate potential-Elastic force Regime.}

\vspace{0.2cm}

In Figures 10(a)-(b)-(c) graphics of both the entanglement entropy
between vacuum-vacuum and kink-kink fluctuations in a $N=500$ particle
chain are shown within the BalReg Regime $g=10^{5}$ for $a=0$, $a=0.6$
and $a=0.99$.

\begin{figure}[h]
\centerline{\includegraphics[width=5cm]{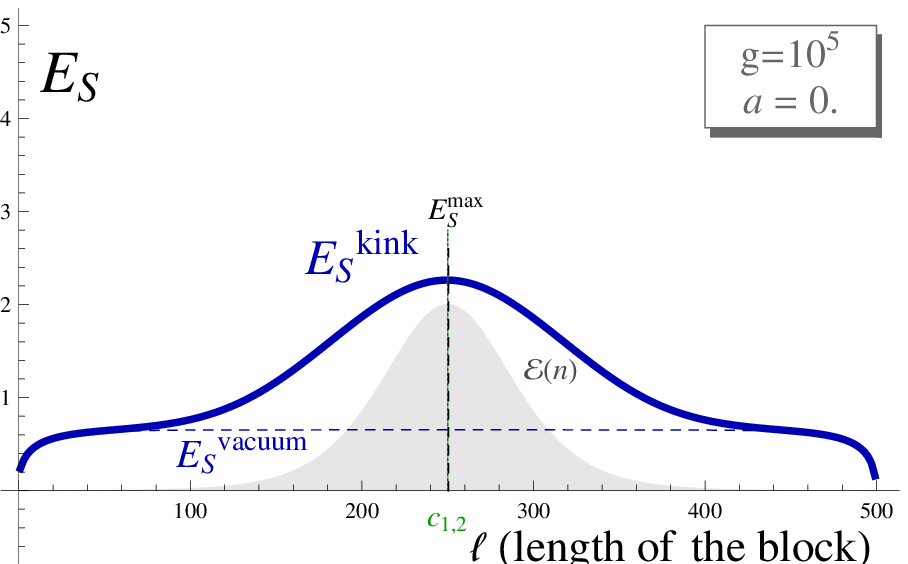}\includegraphics[width=5cm]{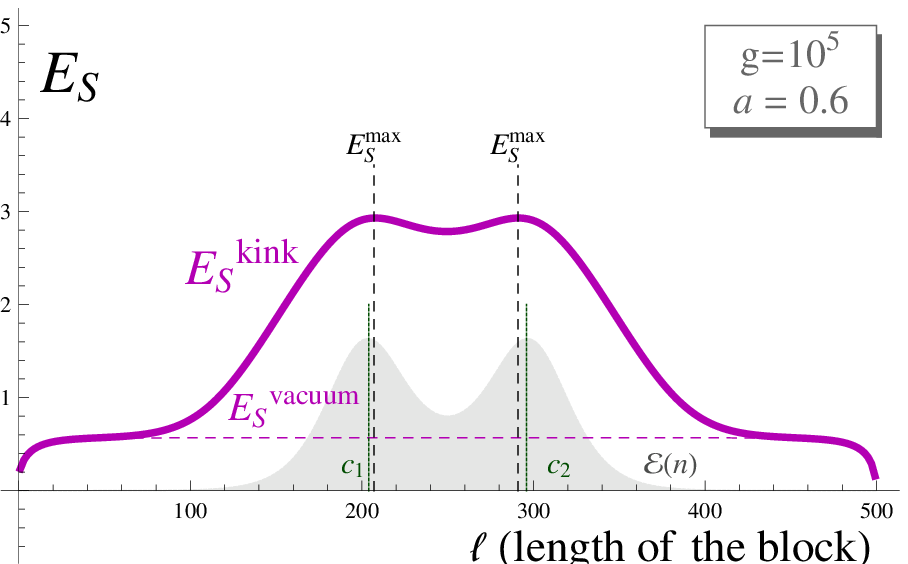}\includegraphics[width=5cm]{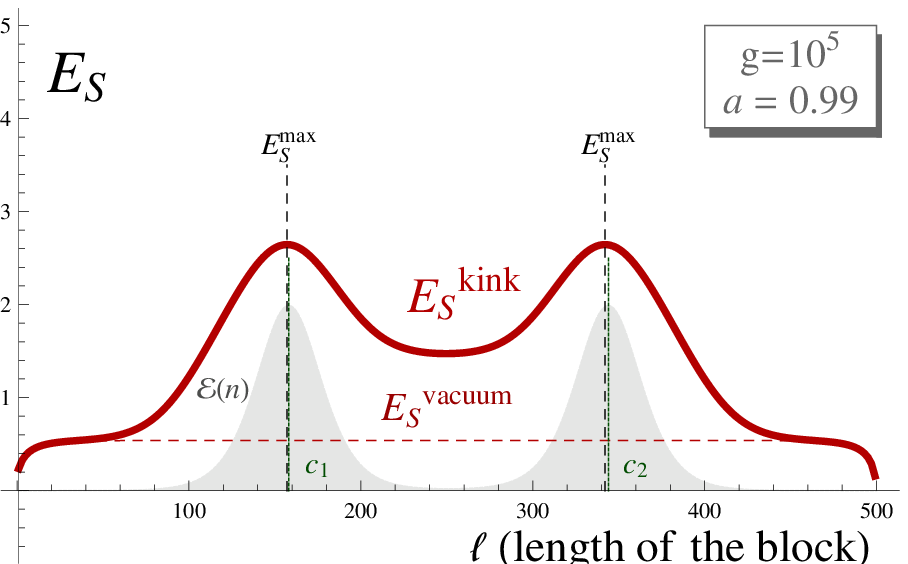}}

\caption{Entanglement entropies between kink-kink (solid line) and vacuum-vacuum
(dashed line) fluctuations for: (a) $a=0$, (b) $a=0.6$ and (c) $a=0.99$
in the BalReg regime. The kink energy distributions are plotted as
curves enclosing a shadowed area using the scale set by $E_{S}$.
Dashed vertical lines show the relative positions of the maxima of
the kink energy distribution and entanglement entropy.}
\end{figure}

\vspace{0.2cm}

There are only quantitative discrepancies with respect to the pattern
just described in the previous $g=10^{4}$ regimes. The curves in
this $g=10^{5}$ regime are smoother and consequently the differences
between the $E_{S}^{{\rm kink}}$ and $E_{S}^{{\rm vacuum}}$ entropies
are less pronounced than those occurring for $g=10^{4}$. The centers
of the kink energy distributions, however, slightly differ from the
maxima of $E_{S}^{{\rm kink}}$ both in the cases when the kink is
formed by one or two lumps. More precisely: (1) For $a=0$ the entropy
$E_{S}^{{\rm kink}}$ shows a maximum at the midpoint of the chain
where the kink center is localized. (2) If $a=0.6$ the function $E_{S}^{{\rm kink}}$
exhibits two maxima situated at $\ell=207$ and $\ell=291$, and,
despite that the respective maxima do not exactly coincide the kink
entanglement entropy and the energy distribution have similar shapes
as functions of $\ell$. (3) For $a=0.9$, the lengths of the subblocks
for which the entanglement entropy is maximum are $\ell=157$ and
$\ell=342$. This means that two subblocks which remarkably differ
in length support maximum entropy. The lump centers nearly coincide
with the entanglement entropy maxima. The little discrepancies between
entropy and kink energy maxima are due to discreteness effects: the
kink energy distribution is more extended in the chain than in the
previous SubsReg regime and the location of the energy distribution
maximum on a discrete chain is less accurate. In Figure 7(b) one observes
that the spatial correlation functions give a clue of the entanglement
entropy shapes also in this regime as one can check in Figures 10(a)-(b)-(c).

\vspace{0.2cm}

C. \textit{Elastic force dominant regime.}

\vspace{0.2cm}

In Figures 11(a)-(b)-(c) the graphics of the entanglement entropies
between kink-kink and vacuum-vacuum fluctuations are plotted in the
$g=10^{6}$ ElasReg Regime for the three usual representative values
of the parameter $a$.

\begin{figure}[h]
\centerline{\includegraphics[width=5cm]{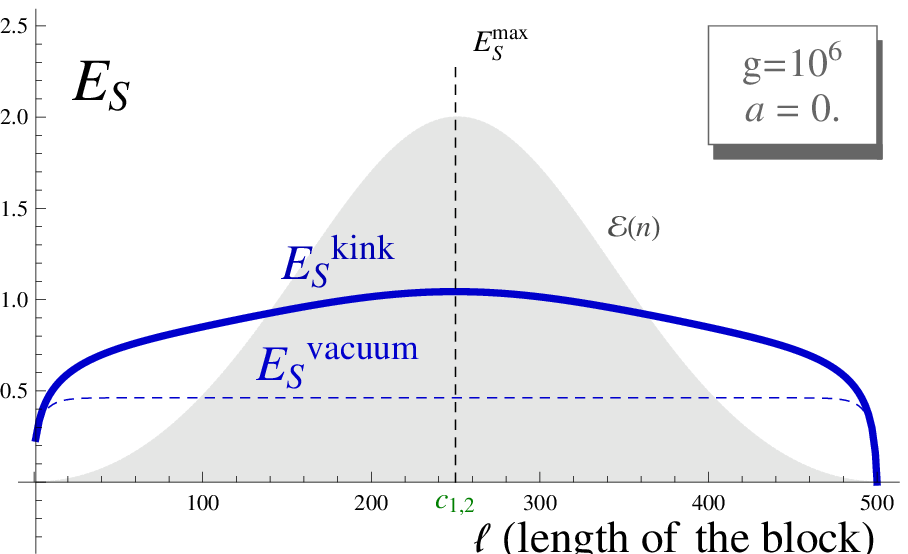}\includegraphics[width=5cm]{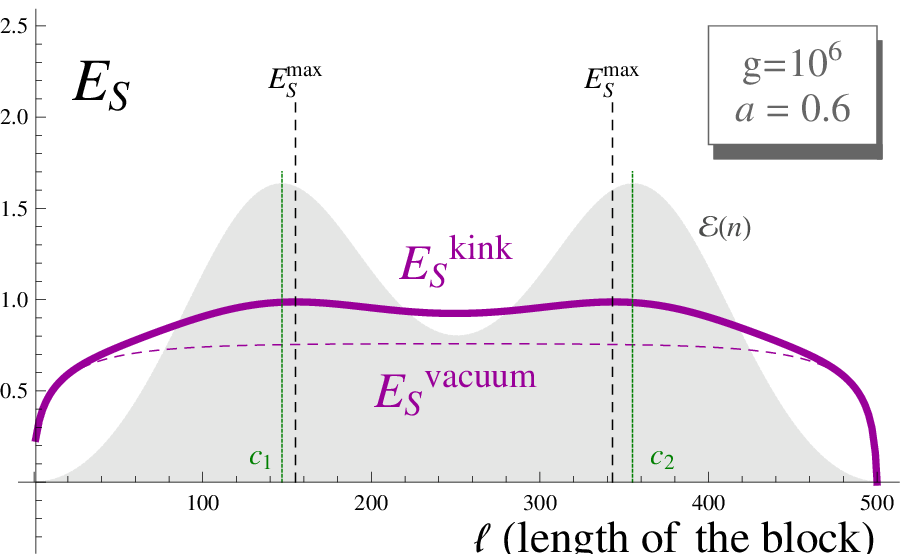}\includegraphics[width=5cm]{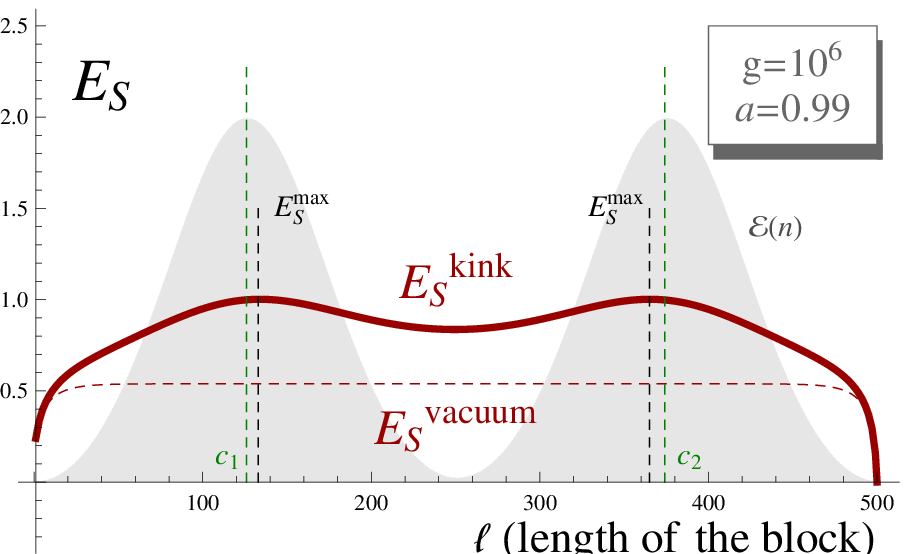}}

\caption{Entanglement entropies between kink-kink (solid line) and vacuum-vacuum
(dashed line) fluctuations for: (a) $a=0$, (b) $a=0.6$ and (c) $a=0.99$
in the ElasReg regime. The kink energy distributions are plotted as
curves enclosing a shadowed area using the same scale as $E_{S}$.
Dashed vertical lines show the relative positions of the kink energy
distribution and the entanglement entropy maxima.}
\end{figure}

\vspace{0.2cm}

Differences between kink-kink and vacuum-vacuum fluctuation entanglement
entropies become small in this regime. In general, these differences
get smaller when the elastic constant $g$ increases and the kinks
become less concentrated. Nevertheless, the shape of the kink entropy
and energy distribution keeps correlated, despite that the center
of the kink energy distribution differs from the maxima of $E_{S}^{{\rm kink}}$
for $a=0.6$ and $a=0.9$ as in the BalReg regime. These discrepancies
are not only due to discreteness effects. Contour effects also affect
because the kink profile differs from a vacuum solution practically
along the whole chain and the boundaries have a stronger influence
on the kink energy distributions. In fact, notice that in this regime
the kink loses its characteristic smooth staircase shape adopting
an almost linear profile, see Figure 6. We remark three points: (1)
The function $E_{S}^{{\rm kink}}$ exhibits a single maximum at the
middle of the chain if $a=0$ as in the other two regimes previously
discussed. (2) For $a=0.6$ the kink fluctuation entropy exhibits
two low curvature maxima at the values $\ell=155$ and $\ell=343$.
(3) The two maxima in the $a=0.99$ case are placed at $\ell=126$
and $\ell=374$. In each case the maxima coincide with that of the
spatial correlation function, Figure 7(c), whereas the entanglement
maxima only nearly coincide with the centers of the lumps of the kink
energy distribution because the reasons mentioned above .

\vspace{0.2cm}

\textbf{Conclusions}. - We have studied entanglement in a chain of
$501$ particles subjected to a substrate potential belonging to the
one-parametric family of double sine-Gordon potentials. We considered
the entanglement arising when a portion of the chain of length $\ell$
is taken from the left of the chain, instead of from the center \cite{Marcovitch08}.
This different approach allowed us to see in a clearer way that, when
focusing on kink solutions carrying an energy distribution either
formed by one or two lumps, the maximum entanglement exhibits a strong
correspondence with both the spatial correlation function and the
centers of the lumps in the kink energy distribution. Regarding the
spatial correlation function, its maxima occur at the same value of
the length of the subblock that makes the entanglement maxima, whereas
the maximum entanglement nearly coincide with the maximum of the lumps
in the kink energy distribution. We conclude that whenever the spatial
correlation function between kink fluctuations presents maxima, also
the kink entanglement entropy and the kink energy distributions exhibit
maxima. In Figure 7 together with Figures 9, 10 and 11 the pattern
previously mentioned is observed. The entanglement grows when the
kink energy distribution evaluated at the sub-block endpoint increases.
In general the maximum for the entanglement is reached when the sub-block
endpoint coincides with the kink lump centers. This pattern is clearly
observed if the kink is confined in a small region, see Figures 9
and 10, where the small discrepancies between kink entanglement entropy
and energy distribution maxima is understood as due to discreteness
effects. In Figure 11 a greater discrepancy between the maxima of
the entanglement and the kink energy distribution is displayed. In
this case the entanglement entropy maximum shift with respect to the
maximum of the energy distribution is mainly associated to boundary
effects. Recall that in this regime the kink occupy essentially all
the chain such that the boundaries influence the kink solution. Indeed
the profile loses its kinky shape, see Figure 6, in favor of an almost
equidistant particle configuration. 
\begin{acknowledgments}
NGA acknowledge financial support from the Brazilian agency CNPq and
thanks Dr. Juan Mateos Guilarte for the kind hospitality during the
stay in USAL. This work was performed as part of the Brazilian National
Institute of Science and Technology (INCT) for Quantum Information. \end{acknowledgments}

\end{document}